\documentclass[aip,preprint]{revtex4-1}
\usepackage{amsmath}
\usepackage{siunitx}
\usepackage{graphicx}
\def\paren#1{\left(#1\right)}
\def\eqref#1{Eq.~\ref{#1}}
\def\figref#1{Fig.~\ref{#1}}
\DeclareMathOperator*{\argmax}{argmax}
\DeclareUnicodeCharacter{2212}{-}
\begin{document}

\title{Microbubble implosions in finite hollow spheres} 

\author{MA H Zosa}
\affiliation{Institute of Laser Engineering, Osaka University, 2-6 Yamadaoka, Suita, Osaka 565-0871, Japan}
\affiliation{Division of Electrical, Electronic and Information Engineering, Graduate School of Engineering, Osaka University, 2-1 Yamadaoka, Suita, Osaka 565-0871, Japan}
\author{M Murakami}
\email[]{murakami-m@ile.osaka-u.ac.jp}
\affiliation{Institute of Laser Engineering, Osaka University, 2-6 Yamadaoka, Suita, Osaka 565-0871, Japan}

\date{\today}

\begin{abstract}
Microbubble implosion (MBI) is a recently proposed novel mechanism with many interesting and exciting potential applications. MBI predicts that the inner layers of a spherical target with a hollow cavity can be compressed into a core with a density of $10^5$ times that of the solid density. Furthermore, this ultra-compressed core mostly consists of ions. This leads to the generation of ultra-high electric fields, which may be applicable to gamma-ray lensing or pair creation. However, MBI has yet to be studied for finite hollow spheres whose electrons are free to redistribute themselves after being given an initial temperature. This paper studies MBI under finite sphere conditions. Using an electron distribution model, the electron distribution after receiving an initial temperature is studied. Then the optimal parameters required to fill a hollow cavity with electrons are calculated. This electron distribution model is used to create a one-dimensional code to simulate the MBI dynamics for finite spheres. MBI occurs even for finite spheres, and high-density compression is still achievable with this setup.
\end{abstract}


\maketitle
\section{Introduction}
The emergence of relativistic laser facilities\cite{Yoon2021,Zhang2020,Haynam2007,Shiraga2011} ($> 10^{18}$ \si{W\per\square\meter}) is a crucial factor in facilitating experiments involving various physical phenomena such as ion acceleration\cite{Torrisi2014,Macchi2010,Henig2009,Bulanov2008,Esirkepov1999}, relativistic electron acceleration\cite{Lu2007,Wagner1997,Arefiev2014,Yu2000,Buck2011}, nuclear fusion\cite{Murakami1995,Murakami2017,Betti2016,Park2014,Ralph2018}, or even quantum electrodynamics (QED)\cite{Burke1997,Titov2018,Krajewska2012,Vranic2018,Gu2018,Gu2018_2,Gu2019}. A more recent proposition, which takes advantage of relativistic laser intensities, is microbubble implosion (MBI)\cite{Murakami2018,Murakami2019}. MBI is a new and exciting proposition because it has various scientific and technological applications such as ion acceleration\cite{Murakami2019}, gamma-ray lensing\cite{Koga2019} and even pair creation\cite{Koga2020}.

In MBI, a target with a spherical cavity is irradiated by an intense laser, which heats the electrons. Once the electrons are heated, they start to fill the cavity and initiate the implosion of ions close to the cavity walls. During maximum compression, the number density, $n$, as a function of $r$ close to the core of the implosion is given by
\begin{equation}
    n(r) = \frac{n_{i0}}{6} \paren{\frac{R_0}{r}}^2
    \label{eq:dens}
\end{equation}
where $n_{i0}$ is the initial ion density and $R_0$ is the initial cavity radius \cite{Murakami2018,Murakami2019}. For the case where $n_{i0} = 5 \times 10^{28} \ \si{\per\cubic\centi\meter}$, the minimum radius is roughly $0.8 \ \si{\nano\meter}$\cite{Murakami2018}. This suggests that the ions are compressed to $10^5$ times that of their original density. This high compression should yield electric fields one order of magnitude lower than the Schwinger limit \cite{Murakami2019}. Hence, vacuum polarization and even pair creation could potentially occur in this scenario \cite{Koga2019,Koga2020}. 

Until now, MBI has been investigated under the assumption that the ion distribution is semi-infinite or periodic or when there are multiple laser beams to ensure a sufficient counterflow of electrons. This would confine most of the electrons within the target, leading to a quasi-neutral plasma. 

This study addresses the case where the electrons are free to redistribute themselves in the presence of a finite ion distribution after they are given an initial temperature. This implies that the target has a net charge because the electrons can escape to infinity \cite{Peano2007}. Although some studies have calculated the electron distribution \cite{Kanapathipillai2004,Peano2007}, they assume either the presence of background ion density or the spherical target is solid. This study does not assume the presence of background ions and includes hollow spherical targets.

\section{Electron distribution model}
To determine the equilibrium electron distribution, Poisson's equation for the electrostatic potential must be solved. It is given as
\begin{equation}
    \nabla^2 \phi = 4 \pi e \paren{n_e - Zn_i} \label{eq:poisson}
\end{equation}
where $e$ is the elementary charge, $Z$ is the ionization state of the ions, $\phi$ is the electrostatic potential, and $n$ is the number density. The subscripts $e$ and $i$ denote electrons and ions, respectively. The standard method of solving $n_e$ assumes that the electrons are isothermal and follow the Boltzmann relation, which is expressed as
\begin{equation}
    n_e = n_{e0} \exp{\paren{e\phi/T}} \label{eq:boltzmann}
\end{equation}
where $T$ is the temperature and $n_{e0}$ is the density as $T \to \infty$. For spherically symmetric cases, \eqref{eq:poisson} and \eqref{eq:boltzmann} can be solved numerically assuming a static ion distribution. However, the Boltzmann relation is problematic when considering finite-sized targets. As noted by Kanapathipillai et al., for finite-sized spherical targets, it is required that $n_e \to 0$ as $r\to \infty$,  which implies that $\phi(\infty)$ diverges \cite{Kanapathipillai2004}. Hence, the expansion of electrons into a vacuum for finite-sized spherical plasmas should not be isothermal. One solution is to add a small background ion density as this ensures that $\phi(\infty)$ does not diverge and \eqref{eq:boltzmann} holds \cite{Kanapathipillai2004}.

To calculate $n_e$ in a vacuum without the background ions, a more generalized form of \eqref{eq:boltzmann} should be used. When the electrons are in hydrostatic equilibrium, an expression for $n_e$ can be obtained by utilizing the polytropic equation of state, $p_e =cn_e^\gamma$, where $p$ is the pressure, $\gamma$ is the polytropic index, and $c$ is a constant \cite{SACK1987311,Manfredi1993}. Thus, $n_e$ can be expressed as
\begin{equation}
    n_e = n_{e0} \paren{1 + \frac{\gamma-1}{\gamma} \frac{e\phi}{T_{e0}}}^{1/\paren{\gamma-1}} \label{eq:ne_poly}
\end{equation}
where the subscript $0$ refers to an unperturbed homogeneous plasma (i.e., $n_{e0} = Zn_{i0}$). Unsurprisingly, \eqref{eq:ne_poly} is equal to \eqref{eq:boltzmann} when considering the isothermal limit ($\gamma \to 1$). Additionally, $\phi(\infty)$ will not diverge under \eqref{eq:ne_poly}.
\section{Electron distribution for a solid sphere}
Using \eqref{eq:poisson} and \eqref{eq:ne_poly}, Poisson's equation for the electric potential of a solid sphere with radius $R_0$ is
\begin{equation}
    \frac{\partial^2 \tilde{\phi}}{\partial \xi^2} + \frac{2}{\xi} \frac{\partial \tilde{\phi}}{\partial \xi} =
    \begin{cases} 
      \Lambda^2 \paren{1 + \frac{\gamma-1}{\gamma}\tilde{\phi}}^{1/\paren{\gamma-1}} - \Lambda^2,  & \xi \leq 1 \\
      \Lambda^2 \paren{1 + \frac{\gamma-1}{\gamma}\tilde{\phi}}^{1/\paren{\gamma-1}}, & \xi > 1 
   \end{cases} 
   \label{eq:poisson_norm}
\end{equation}
where $\xi = r/R_0$, $\tilde{\phi} = e\phi/T$, $\Lambda = R_0 /\lambda_{De} $, and $\lambda_{De}$ is the electron Debye length given by $\sqrt{T_{e0}/\paren{4\pi e^2n_{e0}}}$. Equation \ref{eq:poisson_norm} can be solved using a finite-difference relaxation algorithm.

In the following calculations, the polytropic index, $\gamma$, is $5/3$. The electron densities obtained by solving \eqref{eq:poisson_norm} is compared with three-dimensional molecular dynamics (MD) simulations of electrons. The MD simulations used 10,000 particles, and the ions are modeled as a static uniform cloud of charge. The electrons are initially uniformly distributed in space and given random velocities according to the Maxwell-Boltzmann distribution. The temperatures are calculated to match the desired $\Lambda$. The system is allowed to reach equilibrium, and the time-averaged electron density is calculated.
\begin{figure}[!htbp]
    \centering
    \includegraphics{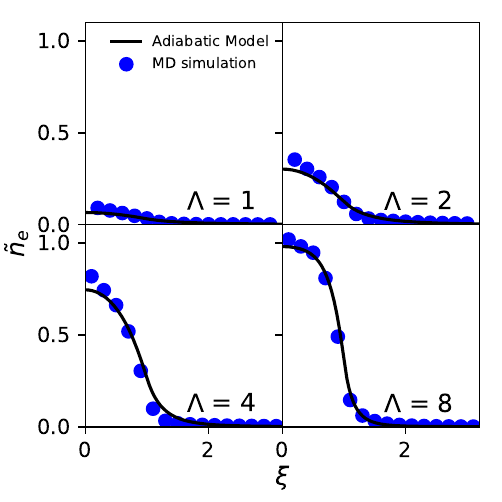}
    \caption{Calculated normalized electron density profile for the model and MD simulations for different values of $\Lambda$.}
    \label{fig:mdsolid}
\end{figure}

Figure \ref{fig:mdsolid} shows the results of $\tilde{n}_e = n_e/n_{e0}$ from the MD simulations and the numerical solution of Poisson's equation. The results agree well for the different values of $\Lambda$. The numerical solution also accurately predicts the known behavior of electrons bound to the ion sphere. As the temperature increases ($\Lambda$ decreases), the number of electrons bound to the ion sphere decreases. When $\lambda_D = R_0$ ($\Lambda = 1$), only a small fraction of electrons are bound to the ion sphere.

\begin{figure}[!htbp]
    \centering
    \includegraphics{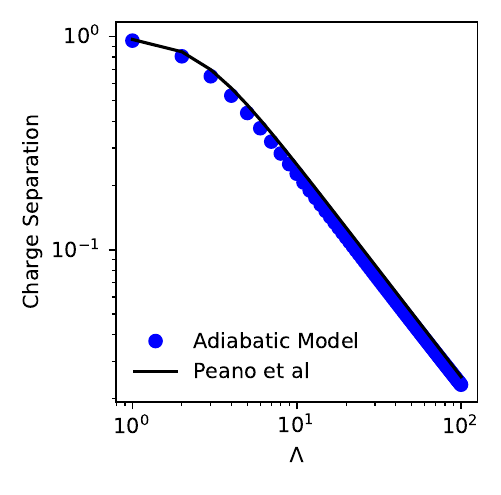}
    \caption{Calculated charge separation from the numerically solved Poisson's equation and the model derived by Peano et al. \cite{Peano2007}}
    \label{fig:peano}
\end{figure}

To further quantify this observation, the charge separation, $\beta$, of the electrons and ion sphere was calculated and compared with the fitted model described by Peano et al.\cite{Peano2007}, which is given by $\beta = F_{2.6}\paren{2.573/\Lambda}$, where $F_\mu(x) = x/(1+x^\mu)^{1/\mu}$. The charge separation $\beta$ is the ratio of the total charge of the electrons within the ion sphere to the ion sphere's total charge. This equation was obtained by fitting the $F_\mu(x) = x/(1+x^\mu)^{1/\mu}$ curve to an electron distribution model \cite{Peano2007}. Our numerical results are almost identical to Peano's fitted model (\figref{fig:peano}). 

\section{Electron occupancy in the cavities of hollow spheres}
For a hollow sphere with an inner radius of $R_{in}$ and an outer radius of $R_{out}$, Poisson's equation for the electric potential can be written as 
\begin{equation}
    \frac{\partial^2 \tilde{\phi}}{\partial \xi^2} + \frac{2}{\xi} \frac{\partial \tilde{\phi}}{\partial \xi} =
    \begin{cases} 
      \Lambda^2 \paren{1 + \frac{\gamma-1}{\gamma}\tilde{\phi}}^{1/\paren{\gamma-1}} - \Lambda^2,  & 1 \leq \xi \leq \xi_{ar} \\
      \Lambda^2 \paren{1 + \frac{\gamma-1}{\gamma}\tilde{\phi}}^{1/\paren{\gamma-1}}, & otherwise
   \end{cases} 
   \label{eq:poisson_hollow}
\end{equation}
where $\xi = r/R_{in}$ and $\xi_{ar} = R_{out}/R_{in}$. The second term on the right hand side, $- \Lambda^2$, corresponds to the ions. It should be noted that all normalizations involving $R_0$ in the previous section are replaced with $R_{in}$ in this section.
\begin{figure}[!htbp]
    \centering
    \includegraphics{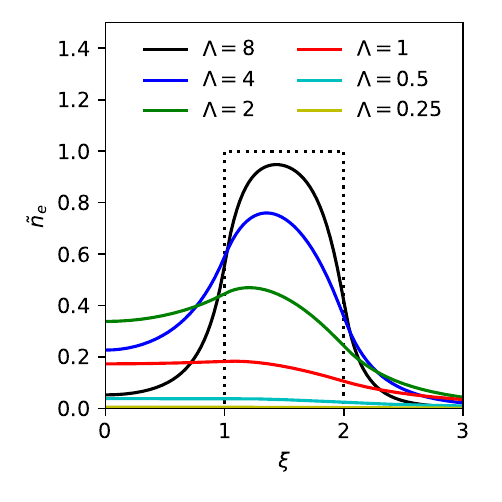}
    \caption{Electron density profile of a hollow target for various values of $\Lambda$. Dotted line denotes the ion density profile.}
    \label{fig:ne_hollow}
\end{figure}

Figure~\ref{fig:ne_hollow} illustrates the electron density profiles. As $\lambda_{De}$ of the electrons increases from $R_{in}/8$ to $R_{in}/2$, the number of electrons filling the cavity increases. This is expected because cold electrons are more likely to stay close to the ion sphere. However, increasing $\lambda_{De}$ past $R_{in}$ decreases the electron distribution because most of the electrons have enough energy to escape the ion sphere completely. This means that, unlike the semi-infinite case where the optimum number of electrons bound in the cavity occurs when $\lambda_{De} \to \infty$ \cite{Murakami2018}, finite spheres have an optimum value of $\Lambda$ where the number of electrons found in the cavity is maximized. Additionally, it can be inferred that the optimum value of $\Lambda$ depends on $\xi_{ar} = R_{out}/R_{in}$ because as $\xi_{ar} \to \infty$, $\Lambda \to 0$ is expected to maximize the number of electrons within the cavity. Therefore, the expression for $\hat\Lambda$ can be written as
\begin{equation}
    \Lambda_{opt}(\xi_{ar}) =\argmax_{\Lambda} \int_0^1 n_e\paren{\xi,\Lambda,\xi_{ar}}\xi^2 d\xi \label{eq:lambda_opt}
\end{equation}
where $\Lambda_{opt}$ is the optimum value of $\Lambda$ for a given $\xi_{ar}$.

\begin{figure}[!htbp]
    \centering
    \includegraphics{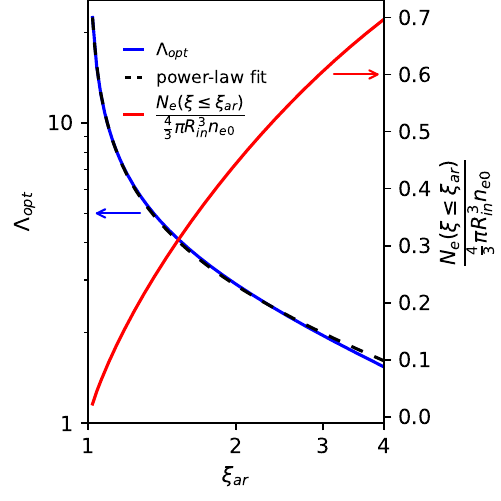}
    \caption{Numerically calculated $\Lambda_{opt}$ as a function of $\xi_{ar}$ along with the power-law fit and the normalized number of electrons within the cavity.}
    \label{fig:lambda_opt}
\end{figure}
Figure~\ref{fig:lambda_opt} shows that as the hollow sphere becomes increasingly thin ($\xi_{ar} \to 1$), $\Lambda_{opt}\to \infty$. Additionally, it shows that $\Lambda_{opt} \to 0$ for $\xi_{ar} \to \infty$. The numerically calculated values for \eqref{eq:lambda_opt} agree well with the power-law fit $\Lambda_{opt} = 2.875\paren{\xi-1}^{-0.5257}$. Additionally, as the targets become thicker, the optimum number of electrons within the cavity, $N_e(\xi \leq \xi_{ar})$, also increases. This is an important factor to consider when designing targets for MBI because the compression ratio of MBI depends largely on the number of electrons within the cavity. Although thinner targets require lower electron temperatures to optimally fill the cavity with electrons, there are fewer electrons within the cavity. This trade-off must be considered when designing spherical MBI targets.
\section{MBI Simulations}
Because our current model does not diverge as $r \to \infty$, it can be utilized to simulate the dynamics of a finite MBI target. To simulate the MBI dynamics of a finite spherical shell, a hybrid one-dimensional spherically symmetric molecular dynamics code was implemented. Similar to previous works \cite{Zosa2020,Murakami2009}, each simulation particle was assumed to be a spherical shell. The protons were treated as spherical shells while the electron density was calculated using the electron model discussed previously. The trajectory of each particle was solved following the equation of motion, which is given by
\begin{equation}
    \frac{dp_{i}}{dt} = \sum_{r_j<r_i} \frac{Q_iQ_j}{r_j^2} + \frac{Q_i^2}{2r_i^2} + Q_iE(r_i)
\end{equation}
where the simulation particles' charge, momentum, and radius are expressed as $Q$, $p$, and $r$, respectively. $E(r)$ is the electric field of the electrons, which can be obtained by solving \eqref{eq:poisson} and \eqref{eq:ne_poly}. Using this model to calculate the electron density and treating the protons as simulation particles, the overall accuracy of the simulation can be maintained while reducing the total number of simulation particles needed because particles to simulate the electrons are not necessary.

For the simulation parameters, the targets were assumed to be composed of pure hydrogen. The default electron temperatures were set to be the optimum temperature as defined by \eqref{eq:lambda_opt}. For all simulations, the cavity's radius was set to \SI{1}{\micro\meter}, and the initial density was set to $n_i= \SI{5d22}{\per\cubic\centi\meter}$. Additionally, for ideal MBI, when the initial density and bubble radius is \SI{5d22}{\per\cubic\centi\meter} and \SI{1}{\micro\meter}, respectively, the bubble will be compressed to a minimum radius of \SI{0.8}{\nano\meter} with a corresponding compression rate of $\sim 10^5$. 

\subsection{Simulation Results}
\begin{figure}[!htbp]
    \centering
    \includegraphics{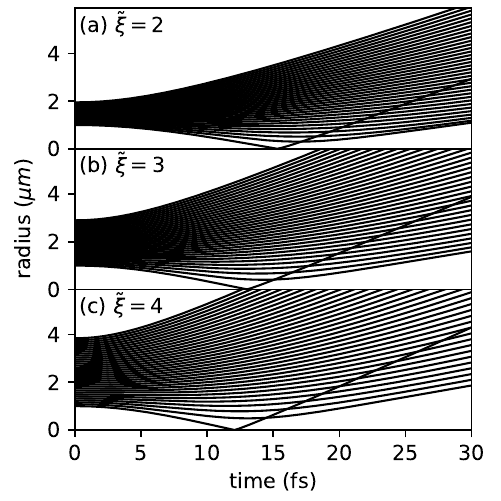}
    \caption{Trajectory of a hollow sphere with \SI{1}{-\micro\meter} cavity undergoing MBI for different aspect ratios: (a) $\xi_{ar} = 2$, (b) $\xi_{ar}=3$, and (c) $\xi_{ar}=4$.}
    \label{fig:trajectory}
\end{figure}

Figure~\ref{fig:trajectory} shows the trajectories of the MBI simulations. Indeed, MBI occurs even for a finite hollow sphere. Furthermore, all three cases considered in \figref{fig:trajectory} indicate proton flashing. Unlike the typical MBI results for semi-infinite geometries, in these cases, charge separation resulting from high electron temperatures causes the the outer layer to expand. Along with the flashed ions from the imploding inner layer, the expanding outer layer is also a source of accelerated ions. As the thickness of the hollow sphere increases, the implosion time decreases (\figref{fig:trajectory}).

MBI is a process that relies heavily on the number of electrons contained within the cavity. As the aspect ratio increases, the maximum number of electrons that can be contained within the cavity also increases (\figref{fig:lambda_opt}). More electrons present within the cavity should increase the implosion velocity of the ions, implying a shorter implosion time \cite{Murakami2018}.

\begin{figure}[!htbp]
    \centering
    \includegraphics{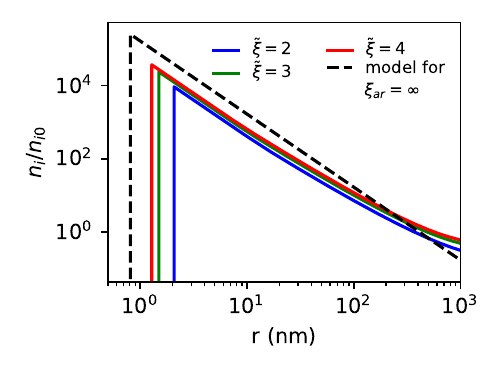}
    \caption{Density profile during maximum compression of a hollow sphere with \SI{1}{-\micro\meter} cavity undergoing MBI for different aspect ratios. The dashed line is the model given by Eq.~\ref{eq:dens}.}
    \label{fig:density}
\end{figure}
All three cases considered can achieve a density compression ratio of $10^4$ times that of the solid density (\figref{fig:density}). This is one order of magnitude lower from the $10^5$ compression ratio predicted using an ideal semi-infinite system given by \eqref{eq:dens}. Nevertheless, \figref{fig:density} illustrates that high compression ratios are possible even with a finite target. More importantly, as the aspect ratio increases, the agreement between the density profile and the theoretical model given by \eqref{eq:dens} improves.

\begin{figure}[!htbp]
    \centering
    \includegraphics{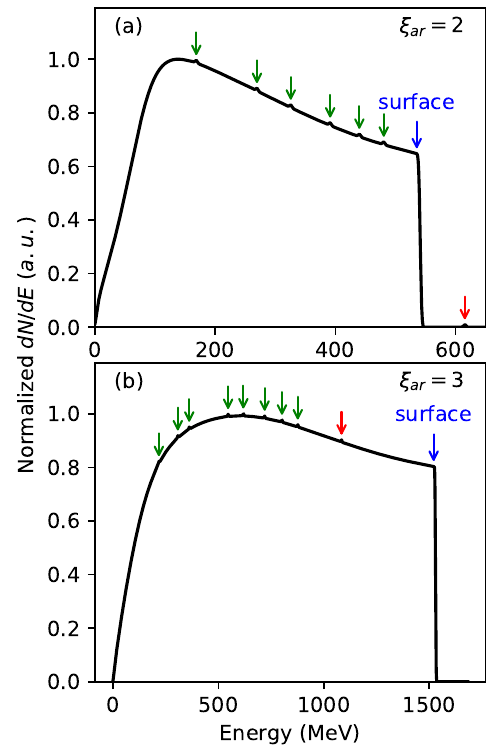}
    \caption{Energy distribution of a hollow sphere with \SI{1}{-\micro\meter} cavity undergoing MBI for: (a) $\xi_{ar} = 2$ and (b) $\xi_{ar}=3$. Green arrows indicate flashed ions, the red arrow indicates the innermost layer, and the blue arrow indicates the surface layer. }
    \label{fig:energy}
\end{figure}
The energy distribution plots show evidence of ion flashing described by earlier studies \cite{Murakami2018,Murakami2019} (\figref{fig:energy}). Comparisons confirm that the small spikes in the graphs agree well with the energies of the inner layers. This implies that these spikes are due to the flashing of the inner ion layers. An interesting observation is that most of the flashed ion layers have lower energies than the outermost coulomb exploding layer. Hence, it is difficult to distinguish the flashed layers from the normal Coulomb exploding layer. However, for $\xi_{ar}=2$, the innermost layer attains an energy of over \SI{600}{\mega\eV}, which is higher than the energy of the outermost coulomb exploding layer. This implies that it would be difficult to detect ion flashing if targets are too thick ($\xi_{ar}>2$).

\begin{figure}[!htbp]
    \centering
    \includegraphics{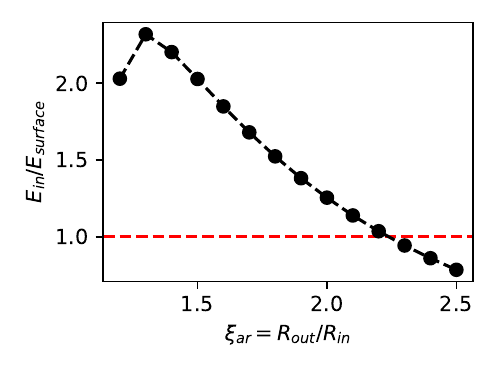}
    \caption{The ratio of the innermost ion energy, $E_{in}$, to the surface ion energy, $E_{surface}$ versus $\xi_{ar}$.}
    \label{fig:Enorm}
\end{figure}
 According to \figref{fig:Enorm}, targets with $\xi_{ar} > 2.2$ will not produce flashed ions. $\Lambda_{opt}$ is larger for thicker targets. As $\Lambda_{opt}$ increases, the charge separation between electrons and ions also increases. This increases the total energy of the coulomb exploding outer layers. On the other hand, the acceleration of the inner layers is influenced by the total number of electrons in the cavity \cite{Murakami2018,Murakami2019}. According to \figref{fig:lambda_opt}, thinner targets have less electrons. This implies that if the target is too thin, the proton flashing is minimal. Figure~\ref{fig:Enorm} shows that when $\xi_{ar} < 1.3$, the energy ratio starts to decrease. Therefore, to maximize proton flashing, the target should have $\xi_{ar} = 1.3$.

\section{Summary and Conclusion}
Using the model described by \eqref{eq:poisson_hollow}, a one-dimensional hybrid code was developed to simulate MBI for finite spheres. MBI occurs for finite spheres. The density compression ratio for the cases studied is $\sim10^4$. For hollow spherical targets with the same cavity radius, thicker targets produce a higher density compression. However, observing ion flashing is more difficult in thicker targets because the coulomb exploding outer layers have much higher energies.

Nevertheless, finite hollow spherical targets are indeed viable for MBI even without containing the electrons within the target. This implies that proof-of-principle experiments can be conducted on finite targets by providing the electrons an initial temperature and allowing the natural time evolution of the system to proceed. This may reduce the complexity required to conduct proof-of-principle experiments about MBI. It should be noted that the target's aspect ratio plays an important role in the different applications of MBI. As we have observed, targets with high aspect ratios should have better compression ratios. Unfortunately, ion flashing in high aspect ratio targets is difficult to detect because the outer layers attain large energies during a Coulomb explosion. Targets with large aspect ratios tend to require more laser energy because their optimal $\Lambda$ is much lower. Although creating a dense core is essential for gamma-ray lensing and pair creation, ion flashing is the essence of ion acceleration applications for MBI. As long as $\xi_{ar}< 2.2$ and $\Lambda_{opt}$ is used, there will be flashed ions. Consequently, finite MBI targets should be tailored to the specific application to be observed.

\begin{acknowledgments}
M. Murakami was supported by the Japan Society for the Promotion of Science (JSPS). M.A.H. Zosa thanks Y. Gu and D. Shokov for fruitful discussions.
\end{acknowledgments}

\section*{Data Availability}
The data that support the findings of this study are available from the corresponding author upon reasonable request.

\section*{Author Declarations}
\subsection*{Conflict of Interest}
The authors have no conflicts to disclose.

\bibliography{references.bib}

\end{document}